\documentclass[12pt]{article}

\usepackage[margin=1in]{geometry}
\usepackage{amsmath,amssymb,amsthm}
\usepackage{mathtools}
\usepackage{authblk}
\usepackage{bm}
\usepackage{enumitem}
\usepackage{natbib}
\usepackage[colorlinks=true,allcolors=blue]{hyperref}
\usepackage{appendix}
\usepackage{setspace}
\usepackage{booktabs}
\usepackage{multirow}
\usepackage{graphicx}
\usepackage{caption}
\usepackage{titlesec}
\usepackage{setspace}
\usepackage{svg}

\titleformat*{\section}{\large\bfseries}

\makeatletter
\def\mojiparline#1{
    \newcounter{mpl}
    \setcounter{mpl}{#1}
    \@tempdima=\linewidth
    \advance\@tempdima by-\value{mpl}zw
    \addtocounter{mpl}{-1}
    \divide\@tempdima by \value{mpl}
    \advance\kanjiskip by\@tempdima
    \advance\parindent by\@tempdima
}
\makeatother

\title{\vspace{-10mm}Covariate-adaptive randomization using historical data and bootstrap-based inference}
\author{$\text{Nodoka Seya}^{1,*}$, $\text{Masataka Taguri}^{1,2}$, $\text{Osamu Togo}^1$ and $\text{Kengo Kamatani}^2$\\
\vspace{-2mm}
\normalsize $^1$ Tokyo Medical University, Tokyo, Japan \\
$^2$ 
The Institute of Statistical Mathematics, Tokyo, Japan\\
$^*$ nodoka.seya.bs@gmail.com
}
\date{}

\newenvironment{keywords}{\par\noindent\textbf{Keywords:}}

\begin{document}

\maketitle

\begin{abstract}
\noindent
Covariate-adaptive randomization (CAR) can improve the efficiency of average treatment effect (ATE) estimation by reducing chance imbalances in prognostic baseline covariates. A recently proposed CAR method that balances multiple covariates, including continuous ones, can achieve optimal precision when the conditional mean of the outcome can be expressed as a linear combination of the covariate features used for the CAR procedure. However, this condition may be difficult to satisfy when the relationship between covariates and the outcome is nonlinear. Furthermore, directly balancing multiple covariates may reduce estimation precision in finite samples. We propose three CAR methods that incorporate a prognostic score learned from individual-level historical data into their imbalance measures. The prognostic score reduces multiple covariates to one dimension while giving larger weights to covariates that are more strongly associated with the outcome. Another challenge with CAR is difficulty in inference due to dependence between subjects induced by the CAR procedure. Existing bootstrap-based inference methods emulate the CAR design by reassigning treatment within each bootstrap sample, which causes difficulty in interval estimation under CAR with continuous covariates. We propose a bootstrap method with unequal probability resampling corresponding to the CAR design, avoiding reassignment within each bootstrap sample. In simulations, the proposed CAR methods reduced prognostic imbalance, thereby improving estimation precision compared to existing assignment designs, and the proposed bootstrap method accurately assessed estimation uncertainty under our proposed CAR. An application to Clazosentan trials demonstrated that the proposed CAR methods similarly improved prognostic balance, and the proposed bootstrap method closely reproduced the distribution induced by the proposed CAR.
\end{abstract}

\begin{keywords}
Causal inference; Covariate-adaptive randomization; Bootstrap; Historical data; Prognostic score; Randomized controlled trial.
\end{keywords}

\section{Introduction}
In randomized controlled trials (RCTs), covariate adjustment can improve efficiency for estimating the average treatment effect (ATE) while guaranteeing consistency \citep{tsiatis2008covadj, lin2013agnostic, wang2023model, zhang2025connection}. 
Covariate adjustment can be conducted at the assignment and/or analysis stage. 
Covariate adjustment at the assignment stage can be performed through covariate-adaptive randomization (CAR).
CAR is a class of random assignment designs that use the assignment and covariate information of subjects assigned before that subject to minimize chance imbalances in covariates between groups.
In contrast, simple randomization (SR) is an unadjusted assignment design that, while guaranteeing that covariate distributions between groups are balanced in expectation, cannot control chance imbalances.
Therefore, many RCTs today adopt CAR rather than SR.
In this study, we present two proposals to address two challenges with CAR.
 
The most common CAR methods, such as stratified randomization and minimization \citep{taves1974minimization, pocock1975mini}, can only account for a few discrete covariates. 
Therefore, if important covariates are continuous or multiple, they must be discretized or selected. 
This may result in important covariates becoming unintentionally imbalanced. 
Although several CAR methods that can account for continuous covariates have been proposed \citep{endo2006minimization, su2011balancing, ma2013balancing, li2019testing}, they are still not derived from some theoretical optimality for estimating ATE.

Recently, \cite{ma2022CAR} proposed a CAR method that can account for multiple covariates, including continuous ones, and proved that even an unadjusted estimator can achieve the minimum asymptotic variance under their CAR method when the conditional mean of the outcome can be expressed as a linear combination of the covariate features used for the CAR procedure.
However, this condition may be difficult to satisfy when the relationship between covariates and the outcome is nonlinear, as is often the case for binary outcomes. 
Furthermore, estimation precision may be worse in finite samples due to the attempt to balance multiple covariates. 
To address these issues, we propose new CAR methods using individual-level historical data (Section \ref{sec: CAR}). 
The key idea is to reduce multiple covariates to a one-dimensional score that captures their association with the outcome.
While using historical data in RCTs has gained attention in recent years, most of this use remains at the analysis stage \citep{pocock1976combination, viele2014use, schuler2022increasing}.

Another challenge with CAR is conducting statistical inference that properly accounts for the assignment mechanism. 
Because each subject’s assignment probability depends on previous assignments and covariates, treatment assignments are dependent across subjects. 
Consequently, conventional analyses that ignore this dependence may yield conservative hypothesis tests and confidence intervals with coverage above the nominal level. 
For example, the usual two-sample $t$-test is conservative under certain CAR procedures \citep{shao2010theory, bugni2018inference}.

Several inference methods for CAR have been proposed, but they have drawbacks.
The model-based analytical approach \citep{ma2022CAR} requires the correct specification of the outcome model.
The model-free analytical approach \citep{ye2022inference}, the two-sample test with adjusted standard errors \citep{bugni2018inference} and the stratified bootstrap method \citep{zhang2020quantile} cannot be used for CAR methods that account for continuous covariates, such as the method of \cite{ma2022CAR} and our proposed methods in Section \ref{sec: CAR}.
The bootstrap test \citep{shao2010theory} cannot be used for constructing confidence intervals.
Existing bootstrap approaches emulate the assignment design by reassigning treatment within each bootstrap sample, which causes difficulties in implementing interval estimation for CAR methods that account for continuous covariates. 
We therefore propose a new bootstrap method that creates the bootstrap distribution corresponding to CAR without reassignment (Section \ref{sec: bootstrap}).
Unlike the existing approaches discussed above, our proposed bootstrap method is designed for a broad range of CAR methods, including those that account for continuous covariates.

The rest of this paper is structured as follows.
In Section \ref{sec: CAR}, we describe our proposed CAR methods using historical data.
In Section \ref{sec: bootstrap}, we describe our proposed bootstrap method that can be used for inference under general CAR methods, including our proposed CAR methods.
In Section \ref{sec: sim}, we conduct simulation studies to evaluate the performance of our proposed methods. 
In Section \ref{sec: app}, we apply our proposed methods to Clazosentan phase 2 and 3 trials \citep{endo2022effects}.
In Section \ref{sec: discuss}, we give concluding remarks.

\section{Covariate-adaptive randomization}
\label{sec: CAR}
In this section, we describe CAR methods.
We briefly review the CAR framework of \cite{ma2022CAR} using only trial data in Section \ref{subsec: CAR ma}, and then propose three CAR methods using historical data in Section \ref{subsec: CAR prop}.
All three methods incorporate prognostic scores learned from historical data into their imbalance measures. The first constructs its imbalance measure within the CAR framework of \cite{ma2022CAR}, whereas the other two use imbalance measures based on the augmentation term of a semiparametric efficient estimator \citep{tsiatis2008covadj}.

For each subject $i$, $X_{i}\in\mathbb{R}^p$ is a $p$-dimensional baseline covariate vector, $A_{i}\in\{0,1\}$ is a treatment assignment (1: treatment, 0: control)
and $Y_{i}\in\mathbb{R}$ is an outcome.
For the random variable $Z$, let the history up to the $j$th subject be $\bar{Z}_{j}:=\{Z_i\}_{i=1}^{j}$.
The trial data $\{X_{i}, A_{i}, Y_{i}\}_{i=1}^{n}$ is fully available for analysis.
However, only $\{\bar{A}_{j-1},\bar{X}_{j}\}$ and the historical data are typically available at the assignment of subject $j$ since the assignment is performed sequentially and outcomes are unknown until the analysis.
The estimand is ATE in the trial.
In RCTs that ensure the causal exchangeability and positivity assumptions, ATE can be expressed as $\theta:=\mathbb{E}[Y\mid A=1]-\mathbb{E}[Y\mid A=0]$ under the causal consistency assumption.

\subsection{A brief review of \cite{ma2022CAR}}
\label{subsec: CAR ma}

\cite{ma2022CAR} proposed the CAR framework with the assignment probability \\
\begin{equation}
    \label{eq: assign prob}
    \mathbb{P}[A_{j}=1\mid \bar{A}_{j-1},\bar{X}_j]=
    \begin{cases}
    \rho & \text{if }\text{Imb}_{j}^{(1)}<\text{Imb}_{j}^{(0)}, \\
    1-\rho                 & \text{if } \text{Imb}_{j}^{(1)}>\text{Imb}_{j}^{(0)},\\
    0.5       & \text{if } \text{Imb}_{j}^{(1)}=\text{Imb}_{j}^{(0)},
  \end{cases}
\end{equation}
where $0.5<\rho <1$ and 
\begin{equation}
   \label{eq: imb}
   \text{{Imb}}_{j}^{(a)}=\left|\left|\sum_{i=1}^{j-1}(2A_{i}-1)\phi(X_{i})+(2a-1)\phi(X_{j})\right|\right|^2,\quad \text{ for }a=0,1.
\end{equation}
$\phi:\mathbb{R}^p \rightarrow \mathbb{R}^q$ is a map from $X_{i}$ to a $q$-dimensional feature space (typically $q>p$), and depending on its specification, it reduces to classical CAR methods such as minimization.
Equation (\ref{eq: imb}) represents the imbalance measure in covariates between groups if subject $j$ were assigned to treatment $a$.
That is, equation (\ref{eq: assign prob}) assigns each subject to the group that reduces covariate imbalance with probability $\rho > 0.5$, given the history of assignments and covariates up to that subject.
For example, if we set $\phi(X)=X$, equation \eqref{eq: assign prob} becomes an assignment algorithm that balances the first moments of covariates.
\cite{ma2022CAR} set $\phi(X)=(1,\sqrt pX^\top,\mathrm{vec}(XX^\top)^\top)^\top$ in simulations, where $\mathrm{vec}(\cdot)$ converts a matrix into a column vector.
We call this method COVCAR.

In the above CAR framework, 
an unadjusted estimator
\begin{equation*}
\hat{\theta}_{n}:=\frac{\sum_{i=1}^{n}A_{i}Y_{i}}{\sum_{i=1}^{n}A_{i}} -\frac{\sum_{i=1}^{n}(1-A_{i})Y_{i}}{\sum_{i=1}^{n}(1-A_{i})}
\end{equation*}
can achieve the minimum asymptotic variance if $\phi(X)$ satisfies
\begin{equation}
\label{eq: optimal}
    g(X):=\frac{\sum_{a=0}^1\{m^{(a)}(X)-\mathbb{E}[m^{(a)}(X)]\}}{2}\in\{\beta^\top\phi(X)\mid \beta \in \mathbb{R}^q\}, 
\end{equation}
where $m^{(a)}(X)=\mathbb{E}[Y\mid A=a,X]$, under the conditions in Theorem 4.4 of \cite{ma2022CAR} regarding the asymptotic normality of $\hat \theta_n$.
We call this property optimal precision, according to \cite{ma2022CAR}.
The function $m^{(0)}(X)$ and $m^{(1)}(X)$ are called prognostic scores, and $g(X)$ is the mean of those scores with centering. 
The fact that the CAR of \cite{ma2022CAR} can achieve optimality when $g(X)$ can be expressed as a linear combination of covariate features used in the CAR procedure corresponds to the fact that ANCOVA, which does not include the product terms of covariates and treatment, can achieve the asymptotic minimum variance under 1:1 SR when $m^{(a)}(X)$ can be expressed as a linear combination of covariates included in the ANCOVA model for $a=0,1$ \citep{tsiatis2008covadj, ye2023CAR}.

\subsection{Proposed covariate-adaptive randomization methods using historical data}
\label{subsec: CAR prop}

Focusing on equation (\ref{eq: optimal}), we see that it is not necessary to balance the multidimensional features $\phi(X)$. 
If we can directly balance $g(X)$ itself, optimal precision can be achieved, even if $m^{(a)}(X)$ cannot be expressed as a linear combination of the elements of $\phi(X)$. 
Moreover, when $m^{(a)}(X)$ can be expressed as a linear combination of the elements of $\phi(X)$, an improvement in the estimation precision of $\hat\theta_n$ can be expected in a finite sample since $g(X)$ reduces the multidimensional covariates $\phi(X)$ to a one-dimensional score in which covariates more strongly associated with the outcome have larger coefficients.
However, since $g(X)$ is unknown in practice and the outcome $Y$ is not available at the assignment phase, it is difficult to train $g(X)$ using trial data. 

Then, we propose CAR methods that train $m^{(a)}(X)$ using historical data and balance $g(X)$.
The detailed procedure is as follows:
\begin{enumerate}
    \item Before the trial: Train a model for the prognostic score $m^{(a)}(x)$ using historical data and denote the trained model as $\widetilde{m}^{(a)}(x)$, for $a = 0,1$. The model used for training need not be a parametric regression model; any machine learning model can be used as long as it satisfies $L_2$-consistency.
    \item Assignment of subject $i$: Use equations \eqref{eq: assign prob} and \eqref{eq: imb} with $\phi(X_{i})=\left(1,\widetilde{g}(X_{i})\right)^\top$, where $\widetilde{g}(X_{i})$ is the estimate of $g(X_i)$ using $\widetilde{m}^{(a)}(x),a=0,1$. 
    The constant 1 is included to balance the sample size.
    $\mathbb{E}[m^{(a)}(X)]$ is estimated by the empirical mean of $\widetilde{m}^{(a)}(X)$ in historical data.
\end{enumerate}
We refer to the proposed method described above as PROCAR1.

We now propose another CAR method that does not require the estimation of $\mathbb{E}[m^{(a)}(X)]$ to improve finite-sample performance under covariate shift.
In SR with sample size $j$, the estimator $\hat \theta_j-\mathrm{aug}_j$ is semiparametric efficient \citep{tsiatis2008covadj}, where
\begin{equation}
\label{eq: aug Aj}
    \text{aug}_j=\sum_{i=1}^{j}\left(A_{i}-\frac{\sum_{i=1}^{j}A_{i}}{j}\right)\left(\frac{\widetilde{m}^{(1)}(X_{i})}{\sum_{i=1}^{j}A_{i}}+\frac{\widetilde{m}^{(0)}(X_{i})}{\sum_{i=1}^{j}(1-A_{i})}\right),
\end{equation}
called an augmentation term.
If the $j$-th subject is assigned to the group $a$ (i.e., $A_j=a$), equation (\ref{eq: aug Aj}) is
\begin{equation*}
\begin{split}
    \text{aug}_j^{(a)}&=\sum_{i=1}^{j-1}\left(A_{i}-\frac{n_{1,j}^{(a)}}{j}\right)\left(\frac{\widetilde{m}^{(1)}(X_{i})}{n_{1,j}^{(a)}}+\frac{\widetilde{m}^{(0)}(X_{i})}{n_{0,j}^{(a)}}\right)\\
    &\quad +\left(a-\frac{n_{1,j}^{(a)}}{j}\right)\left(\frac{\widetilde{m}^{(1)}(X_{j})}{n_{1,j}^{(a)}}+\frac{\widetilde{m}^{(0)}(X_{j})}{n_{0,j}^{(a)}}\right),
\end{split}
\end{equation*}
where $n_{1,j}^{(a)}=\sum_{i=1}^{j-1}A_{i}+a$ and $n_{0,j}^{(a)}=j-n_{1,j}^{(a)}$. 
For notational convenience, we define $\text{aug}_j^{(a)}=0$ if $n_{1,j}^{(a)}=0$ or $n_{0,j}^{(a)}=0$.
Equation \eqref{eq: imb} with $\phi(X_{i})=\left(1,\widetilde{g}(X_{i})\right)^\top$ equals zero if and only if $n_{1,j}^{(a)}=n_{0,j}^{(a)}={j}/{2}$ and $\text{aug}_j^{(a)}=0$ hold.
Thus, PROCAR1 can be intuitively interpreted as being designed so that the sample sizes across groups become equal and that the augmentation term of the semiparametric efficient estimator becomes zero (i.e., the unadjusted estimator $\hat{\theta}_j$ is semiparametric efficient).
Based on this, we propose setting the imbalance measure as 
\begin{equation}
\label{eq: imb aug}
  \text{Imb}_j^{(a)}=\left\{n_{1,j}^{(a)}-n_{0,j}^{(a)}\right\}^2+\mathbb{I}\left(n_{1,j}^{(a)}n_{0,j}^{(a)}\neq0\right)\left\{j\times\text{aug}_j^{(a)}\right\}^2,
\end{equation}
instead of equation \eqref{eq: imb}.
We refer to this method as PROCAR2.

We also propose a CAR method that evaluates the imbalances in sample size and prognostic score separately, rather than combining them, to prevent a decline in performance due to scale differences between the first and second terms in equation \eqref{eq: imb aug}.
Specifically, we propose the CAR method with the assignment probability set to
\begin{equation}
    \label{eq: assign prob 2step}
    \begin{split}
    &\mathbb{P}[A_{j}=1\mid \bar{A}_{j-1},\bar{X}_j]\\
    &=
    \begin{cases}
    \rho & \text{if }\sum_{i=1}^{j-1}(2A_{i}-1)<-\delta\text{ or }\left\{|\sum_{i=1}^{j-1}(2A_{i}-1)|\leq \delta\text{ and }|\text{aug}_{j}^{(1)}|<|\text{aug}_j^{(0)}|\right\}, \\
    1-\rho                 & \text{if }\sum_{i=1}^{j-1}(2A_{i}-1)>\delta \text{ or }\left\{|\sum_{i=1}^{j-1}(2A_{i}-1)|\leq \delta\text{ and }|\text{aug}_j^{(1)}|>|\text{aug}_j^{(0)}|\right\},\\
    0.5       & \text{if } |\sum_{i=1}^{j-1}(2A_{i}-1)|\leq \delta\text{ and }|\text{aug}_j^{(1)}|=|\text{aug}_j^{(0)}|,
  \end{cases}
    \end{split}
\end{equation}
where $\delta$ is the pre-specified tolerance value for imbalance in sample size.
Equation \eqref{eq: assign prob 2step} implies assigning the subject to the group with the smaller sample size with probability $\rho$ if the difference in sample sizes between groups is greater than $\delta$ (without accounting for $|\text{aug}_j^{(a)}|$); and assigning the subject to the group $a$ where $|\text{aug}_j^{(a)}|$ is smaller with probability $\rho$ if the difference in sample sizes between the groups is less than or equal to $\delta$ (without accounting for the imbalance in sample sizes).
We refer to this method as PROCAR3.

All proposed CAR methods are designed to minimize the imbalance in $\widetilde g(X)$ and guarantee consistency for ATE regardless of model misspecification or distribution shift.
Model misspecification or distribution shift affects how much the estimation precision can be improved.
The proposed CAR methods differ in finite samples.
Since PROCAR2 and PROCAR3 do not require the estimation of $\mathbb{E}[m^{(a)}(X)]$, they may improve finite-sample performance under covariate shift compared to PROCAR1.
If the historical data include only control subjects, we set $\widetilde{m}^{(1)}(x)$ as $\widetilde{m}^{(0)}(x)$. 
In this case, achieving optimal precision requires the homogeneous treatment effect assumption, i.e., ${m}^{(1)}(x)-{m}^{(0)}(x)=\theta$.

\section{Bootstrap-based inference under covariate-adaptive randomization}
\label{sec: bootstrap} 
In this section, we propose a bootstrap method applicable to inference under general CAR methods, including our CAR methods proposed in Section \ref{sec: CAR}.
We describe the key idea in Section \ref{subsec: bootstrap idea}, and the detailed procedure of our proposed bootstrap method in Section \ref{subsec: bootstrap method}.

\subsection{Idea for proposed bootstrap method}
\label{subsec: bootstrap idea}
We denote the bootstrap sample as $\{X_{b,i}^*,A_{b,i}^*,Y_{b,i}^*\}_{i=1}^n$ for $b=1,...,B$, where $B$ is the number of resampling iterations.
The likelihood of a naive bootstrap sample obtained by simply performing random resampling at equal probabilities of $1/n$ with replacement from the original sample is \\
\begin{equation}
\label{eq: naive likelihood}
\prod_{i=1}^{n}f_{n}(Y^*_{b,i}\mid A_{b,i}^*,X_{b,i}^*)f_{n}(A_{b,i}^*\mid {X}_{b,i}^*)f_{n}(X_{b,i}^*),
\end{equation}
where $f_{n}$ denotes an empirical distribution.
Equation (\ref{eq: naive likelihood}) is simply a decomposition of i.i.d. data.
However, as is obvious from equation (\ref{eq: assign prob}), because the design makes a subject’s assignment probability dependent on the assignments of previous subjects and covariates, the data are not independent under CAR.

The (ideal) likelihood corresponding to the assignment design is 
\begin{equation}
\label{eq: ideal likelihood}
\prod_{i=1}^{n}f_{n}(Y^*_{b,i}\mid A_{b,i}^*,X_{b,i}^*)f(A_{b,i}^*\mid \bar{A}_{b,i-1}^*,\bar{X}_{b,i}^*)f_{n}(X_{b,i}^*),
\end{equation}
where $f(A_{b,i}^*\mid \bar{A}_{b,i-1}^*,\bar{X}_{b,i}^*)$ is the conditional distribution of the assignment corresponding to CAR.
Therefore, if we can resample with \textit{unequal probabilities} proportional to 
\begin{equation*}
\frac{f(A_{b,i}^*\mid \bar{A}_{b,i-1}^*,\bar{X}_{b,i}^*)}{f_n(A_{b,i}^*\mid {X}_{b,i}^*)},
\end{equation*}
which is the ratio of the $i$-th contribution in equation (\ref{eq: ideal likelihood}) to that in (\ref{eq: naive likelihood}), we should be able to obtain a bootstrap sample corresponding to equation (\ref{eq: ideal likelihood}).

\subsection{Proposed bootstrap method}
\label{subsec: bootstrap method}
Based on the idea in Section \ref{subsec: bootstrap idea}, we propose the bootstrap method.
For each bootstrap sample $b=1,...,B$, we sequentially resample $n$ subjects, where $i=1,...,n$ denotes the resampling step and $j=1,...,n$ indexes subjects in the original sample. At the $i$-th resampling step:
\begin{enumerate}
    \item Calculation of assignment probabilities: For each subject $j$ in the original sample, we calculate $f(A_j\mid\bar{A}_{b,i-1}^*,\bar{X}_{b,i-1}^*,X_j)$. Informally, a large probability $\rho$ is given to subjects with a combination of $\{X_j,A_j\}$ that reduces imbalance, while a small probability $1-\rho$ is given to subjects with a combination of $\{X_j,A_j\}$ that increases imbalance.
    \item Unequal-probability resampling: Resample one subject with replacement from $n$ subjects of the original sample, with a probability proportional to $ f(A_j\mid\bar{A}_{b,i-1}^*,\bar{X}_{b,i-1}^*,X_j)/f_n(A_j\mid X_j)$, and use the resampled data as $\{X_{b,i}^*,A_{b,i}^*,Y_{b,i}^*\}$. In the case of our proposed CAR dealing with continuous prognostic scores, $f_n(A_j\mid X_j)\propto 1$.
\end{enumerate}
In summary, we construct a bootstrap sample corresponding to CAR by making subjects whose covariate–treatment combinations reduce the imbalance measure more likely to be resampled. 

Let $\mathcal{B}:=\left\{b\in \{1,..., B\} : 0 < \sum_{i=1}^{n} A_{b,i}^{*} < n\right\}$ be the set of bootstrap samples that include at least one subject from each group.
For $b\in \mathcal{B}$, calculate
\begin{equation*}
\hat\theta_{b,n}^*:=
        \frac{\sum_{i=1}^nA_{b,i}^*Y_{b,i}^*}{\sum_{i=1}^nA_{b,i}^*}-\frac{\sum_{i=1}^n(1-A_{b,i}^*)Y_{b,i}^*}{\sum_{i=1}^n(1-A_{b,i}^*)}.
\end{equation*}
Then, the variance is estimated as follows:
\begin{equation*}
     \widehat{\mathbb{V}}[\hat\theta_n]:=\frac{1}{|\mathcal{B}|}\sum_{b \in \mathcal{B}}\left(\hat\theta_{b,n}^*-\frac{1}{|\mathcal{B}|}\sum_{b \in \mathcal{B}}\hat\theta_{b,n}^*\right)^2.
\end{equation*}
The $100(1-\alpha)$ percent confidence interval is constructed based on a $t$-distribution approximation as $\left[\hat\theta_n-t_{\alpha/2}(n-2)\sqrt{\widehat{\mathbb{V}}[\hat\theta_n]},\quad \hat\theta_n+t_{\alpha/2}(n-2)\sqrt{\widehat{\mathbb{V}}[\hat\theta_n]}\right]$, where $t_{\alpha/2}(n-2)$ is the $100(1-\alpha/2)$-th percentile of the $t$-distribution with $n-2$ degrees of freedom. 
The key feature of our proposed method lies in the construction of the bootstrap sample, and the subsequent procedures are the same as those in the standard bootstrap method.

\section{Simulation studies}
\label{sec: sim}
In this section, we conduct simulation studies to evaluate the performance of our proposed CAR methods and our proposed bootstrap method.

Trial data sizes are varied as $n=20, 50, 100, 200, 500$.
For trial data, covariates $X=(X_{1},...,X_{5})^\top$ are generated as $X_{1}\sim Bin(1,\alpha_1)$, $X_{2}\mid X_{1}\sim Bin(1,\mathrm{expit}(0.5+\alpha_2X_{1}))$ and $(X_{3},X_{4},X_{5})^\top \sim MVN((\alpha_3,\alpha_4,\alpha_5)^\top,V)$ with $(\alpha_1,\alpha_2,\alpha_3,\alpha_4,\alpha_5)^\top=(0.5,0.5,5,50,120)^\top$.
By denoting the $(i,j)$-th element of $V$ as $v_{i,j}$, $v_{1,1}=v_{3,3}=4$, $v_{2,
2}=25$, $v_{1,2}=v_{2,1}=v_{2,3}=v_{3,2}=-4$ and $v_{1,3}=v_{3,1}=1.6$.
The generation of the treatment assignment depends on the assignment method. 
The outcome is generated as $Y\mid A, X\sim N(\mathbb{E}[Y\mid A,X],20^2)$, where $\mathbb{E}[Y \mid A, X]
= 100 + 0.1 (
-0.05 X_1 + 0.8 X_2 + 1.2 X_3 + 0.8 X_4 - 0.1 X_5
+ 4 X_1 X_2 + 2 X_1 X_3 + X_1 X_4 + 2 X_1 X_5
+ 0.1 X_2 X_3 - 0.2 X_3 X_4 + 0.1 X_4 X_5
- 0.05 X_3^2 + 0.05 X_4^2 - 0.05 X_5^2)$.
Thus, $m^{(1)}(X)=m^{(0)}(X)$ and $\theta=\mathbb{E}[Y\mid A=1]-\mathbb{E}[Y\mid A=0]=0$.
For historical data, the sample size is $10000$, and data generation is the same as trial data, except for $A\equiv0$ and $(\alpha_1,\alpha_2,\alpha_3,\alpha_4,\alpha_5)^\top=(0.3,0.3,0,40,100)^\top$.
That is, historical data consists solely of control subjects, and there is a covariate shift.
The simulation is repeated 500 times.

The following five assignment designs are compared: SR, COVCAR \citep{ma2022CAR}, PROCAR1, PROCAR2 and PROCAR3 with $\delta=2$.
In COVCAR, $\phi(X)$ is set as $\phi(X)=(1,\sqrt 5X^\top,\mathrm{vec}(XX^\top)^\top)^\top$.
In PROCAR1--3, the model for $m^{(0)}(x)$ is correctly specified. 
All four CAR methods (COVCAR and PROCAR1--3) can achieve optimal precision with these settings.
For any of the five assignment methods, we use an unadjusted estimator $\hat \theta_n$ in the analysis stage.
Additionally, we include $\hat \theta_n -\mathrm{aug}_n$ which is a semiparametric efficient estimator \citep{tsiatis2008covadj} using historical data to train the prognostic score under SR (proaipw) for comparison with covariate adjustment at the analysis stage.

\begin{figure}[h]
\centering
    \includegraphics[width=\linewidth]{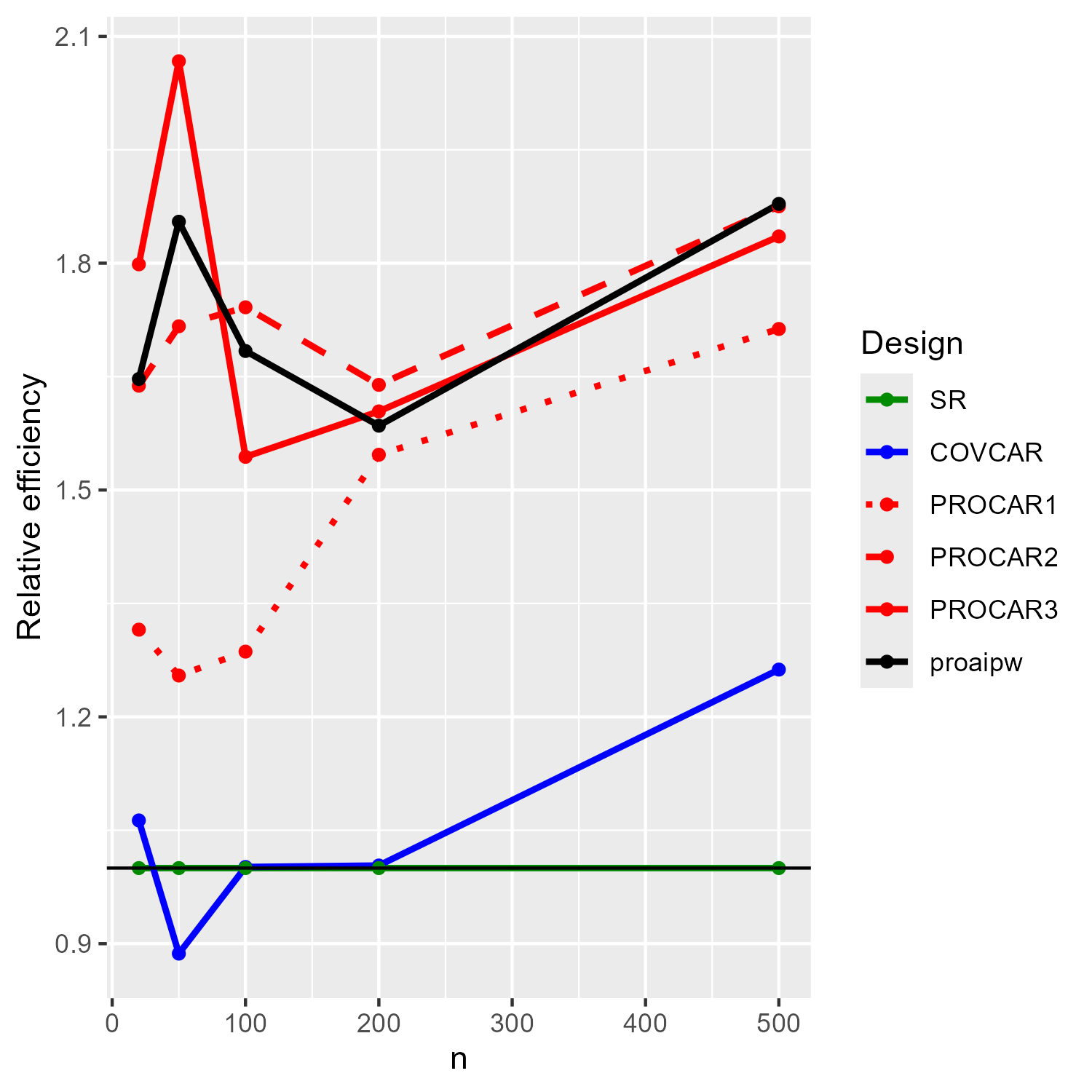}
    \caption{Simulation results for the assignment designs combined with estimator $\hat \theta_n$ at the analysis stage in terms of estimation precision for $\theta$. The x-axis represents the sample size. The y-axis represents the relative efficiency which is the variance of $\hat{\theta}_n$ under simple randomization (SR) over 500 iterations divided by the variance of $\hat{\theta}_n$ under each assignment design: SR (green), COVCAR (\cite{ma2022CAR}; blue), PROCAR1 (red dotted), PROCAR2 (red dashed), PROCAR3 (red solid), over 500 iterations. The black line represents the variance of $\hat{\theta}_n$ under SR over 500 iterations divided by the variance of proaipw $\hat{\theta}_n-\mathrm{aug}_n$ under SR. The black horizontal line represents relative efficiency of 1.}
    \label{fig:a}
\end{figure}

\begin{figure}
    \centering
    \includegraphics[width=\linewidth]{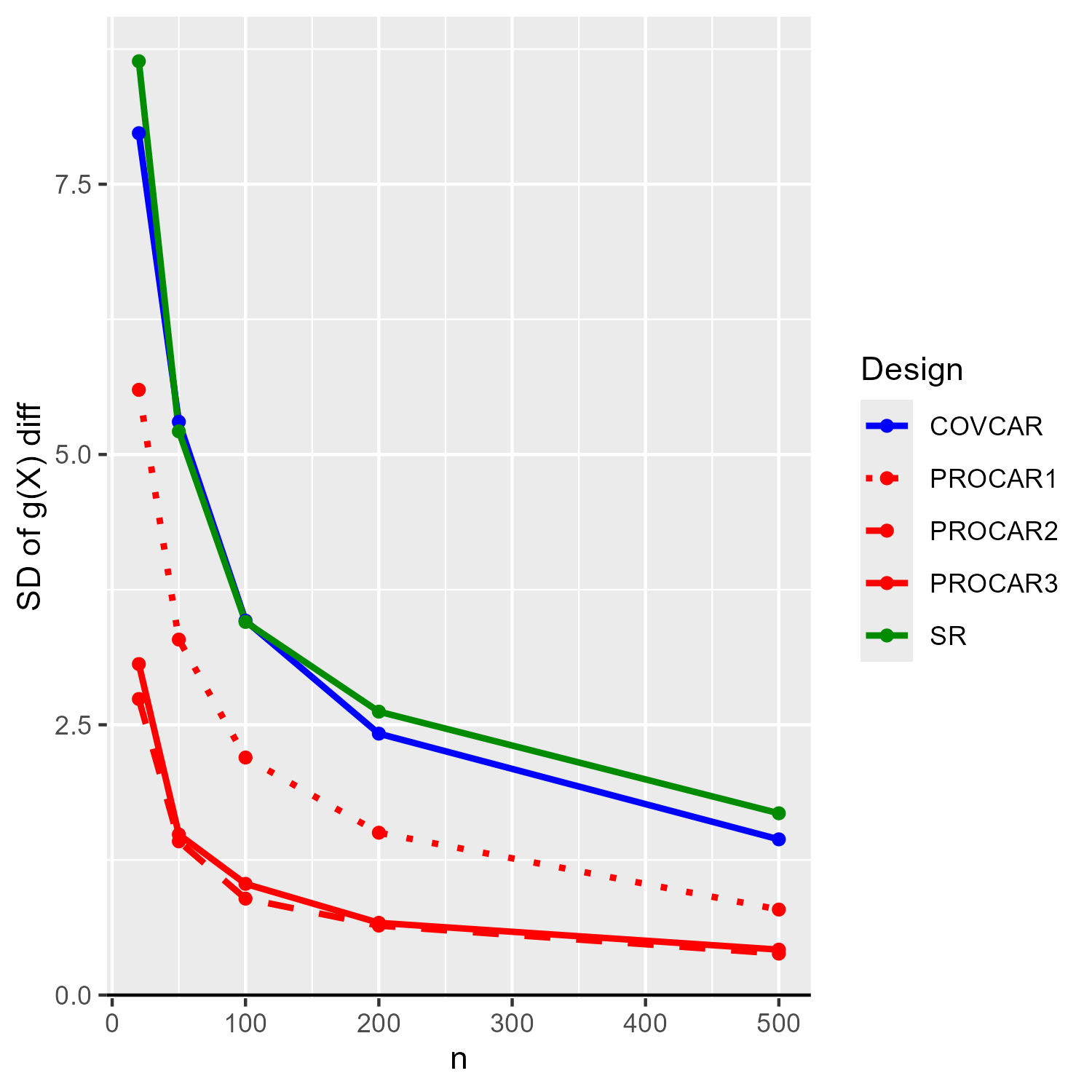}
     \caption{Simulation results for the assignment designs in terms of the balance of $g(X)$. The x-axis represents the sample size. The y-axis represents the standard deviation of the between-group difference in the mean of ${g}(X)$ (i.e., $\sum_{i=1}^n A_i {g}(X_i)/\sum_{i=1}^n A_i - \sum_{i=1}^n (1-A_i){g}(X_i)/\sum_{i=1}^n(1-A_i)$) over 500 iterations. The red lines represent proposed CAR methods; the dotted line represents PROCAR1, the dashed line represents PROCAR2, the solid line represents PROCAR3. The blue line represents COVCAR \citep{ma2022CAR} and the green line represents simple randomization (SR).}
    \label{fig:b}
\end{figure}

Figures \ref{fig:a} and \ref{fig:b} show simulation results for the assignment designs.
For both relative efficiency and $g(X)$ balance, 
PROCAR1--3 show better performance compared to existing assignment designs.
In terms of relative efficiency, despite using an unadjusted estimator at the analysis stage, PROCAR1--3 achieve efficiency comparable to that of the semiparametric efficient estimator, with PROCAR3 showing slightly better performance in small samples.

\begin{figure}
    \centering
    \includegraphics[width=\linewidth]{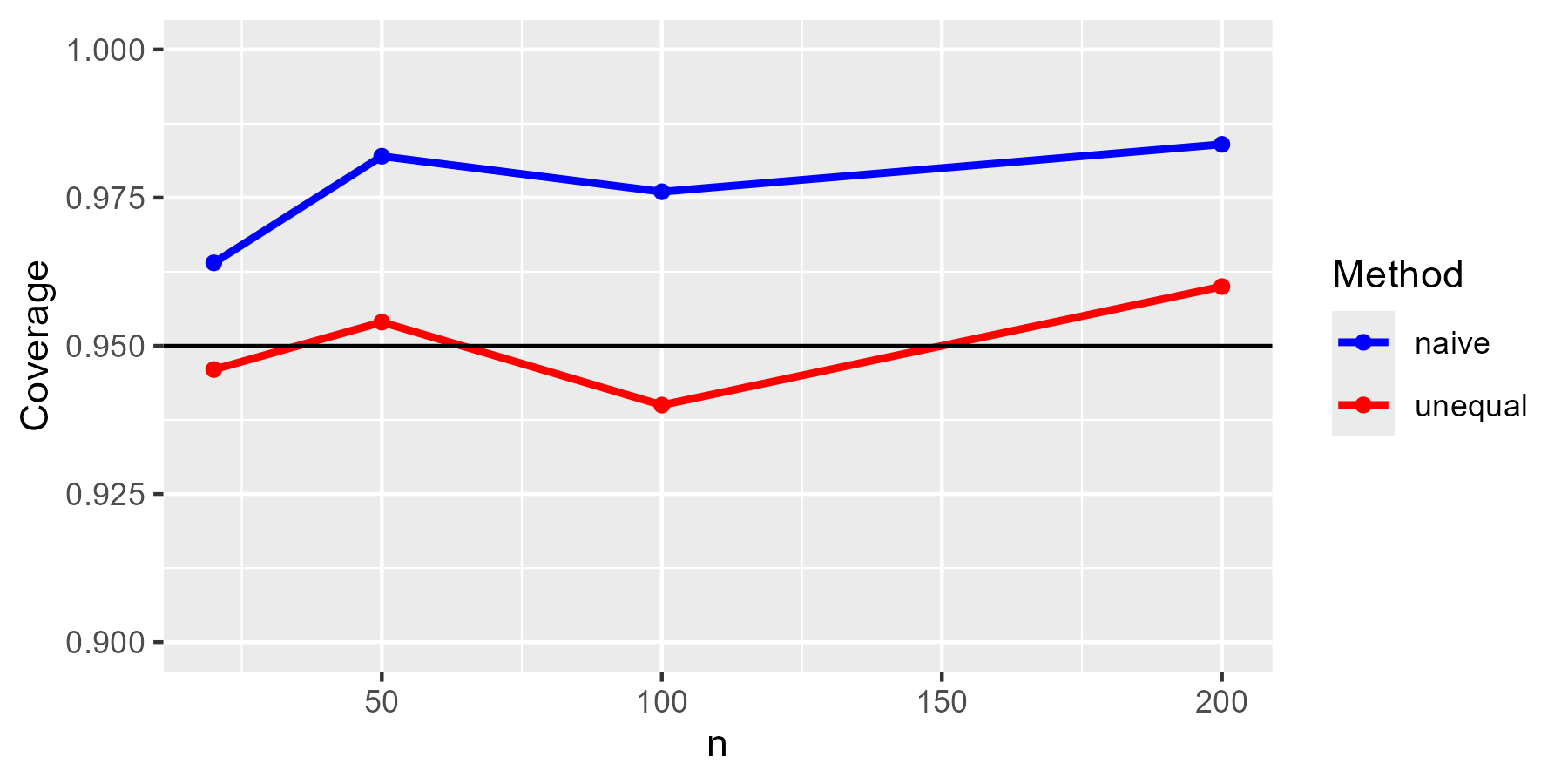}
    \caption{Simulation results for the bootstrap methods.  The x-axis represents the sample size. The y-axis represents the coverage probability of the 95 \% confidence interval under PROCAR1. The red line represents our proposed bootstrap method (unequal). The blue line represents naive bootstrap method with resampling probability $1/n$ (naive). The black horizontal line represents coverage probability of $0.95$.}
    \label{fig:c}
\end{figure}

We next evaluate the performance of the proposed bootstrap method for interval estimation.
The following two inference methods are compared: naive bootstrap method with resampling probabilities $1/n$ (naive) and our proposed bootstrap method with unequal resampling probabilities (unequal). 
For both bootstrap methods, we evaluate the performance of interval estimation under PROCAR1 with $B=500$.
Figure \ref{fig:c} shows simulation results for the two bootstrap methods.
The naive bootstrap method overestimates variance and becomes conservative, while our proposed bootstrap method approximately achieves the nominal coverage probability.
The mean confidence interval width ratio (naive/unequal) is about 1.2 for all sample sizes, and our proposed bootstrap method yields narrower confidence intervals.

\section{Data application}
\label{sec: app}
In this section, we apply our proposed CAR methods and bootstrap method to Clazosentan phase 2 and 3 trials \citep{endo2022effects}.

Clazosentan phase 2 trial is a multi-regional randomized phase 2 trial to evaluate the efficacy and safety of clazosentan for the prevention of cerebrovascular spasm for patients who underwent clipping surgery for aneurysmal subarachnoid hemorrhage.
The original assignment design of Clazosentan phase 2 trial is minimization with WFNS grade in 2 categories (I, II / III, IV) as an assignment factor, where subjects were randomly assigned to either the clazosentan 5 mg/hr group, the clazosentan 10 mg/hr group or the placebo group.
Clazosentan phase 3 trial is a randomized phase 3 trial in Japan to evaluate the effect of clazosentan on cerebral vasospasm-related morbidity and all-cause mortality for patients who underwent clipping surgery for aneurysmal subarachnoid hemorrhage.
The original assignment design of Clazosentan phase 3 trial is minimization with WFNS grade in 4 categories (I / II / III / IV) as an assignment factor, where subjects were randomly assigned to either the clazosentan 10 mg/hr group or the placebo group.

\begin{figure}
    \centering
    \includegraphics[width=\linewidth]{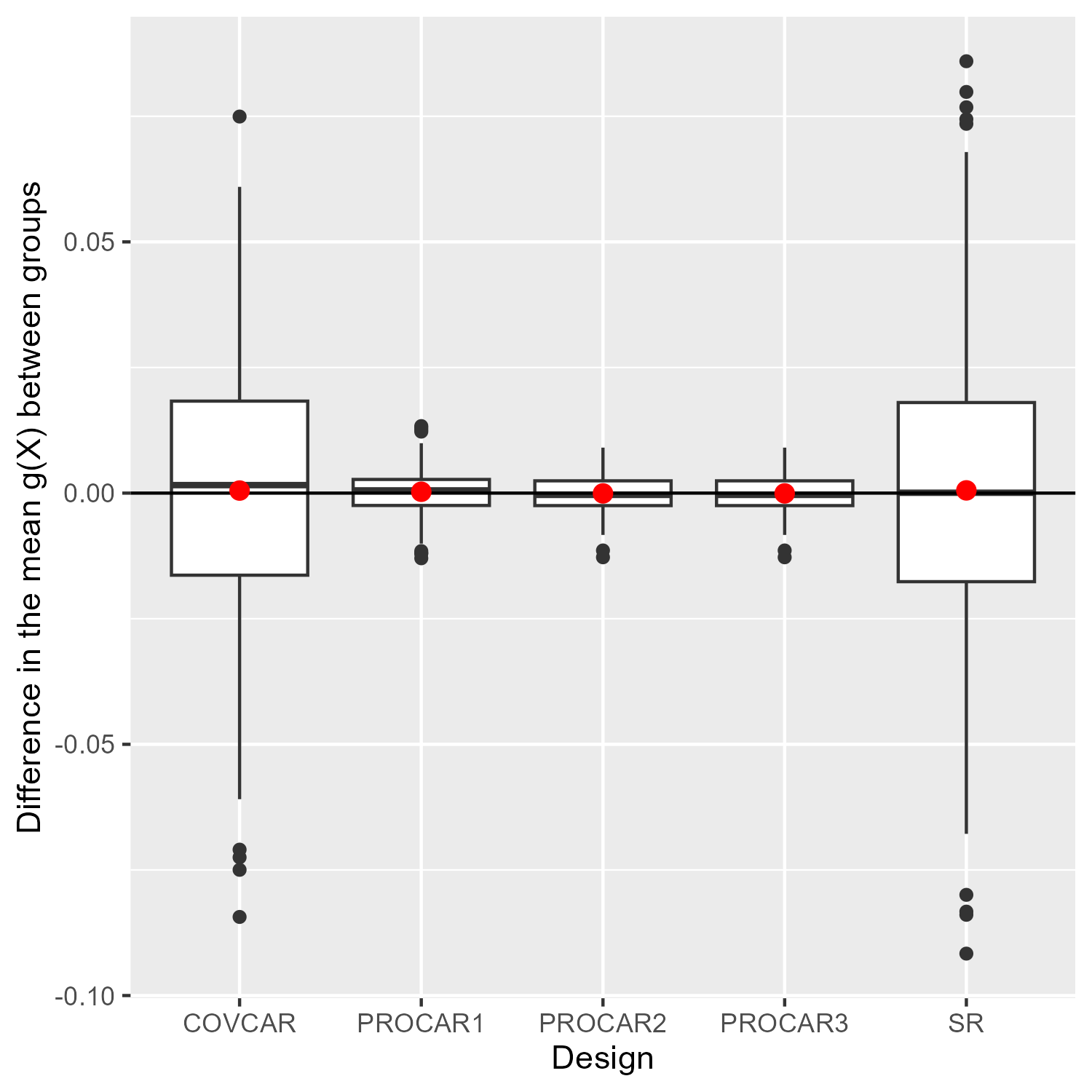}
    \caption{Results of applying the assignment designs to Clazosentan phase 3 trial data. Boxplot of the between-group difference in the mean of $\widetilde{g}(X)$ (i.e., $\sum_{i=1}^n A_i \widetilde{g}(X_i)/\sum_{i=1}^n A_i - \sum_{i=1}^n (1-A_i)\widetilde{g}(X_i)/\sum_{i=1}^n(1-A_i)$) by changing the random seed and repeating the assignment for 500 iterations. The red dots represent the mean of the between-group difference in the mean of $\widetilde{g}(X)$ over 500 iterations. The black horizontal line represents a between-group difference of zero.}
    \label{fig: ana prog plot}
\end{figure}

We use the clazosentan 10 mg/hr group and the placebo group of Clazosentan phase 2 trial as historical data (the sample size is 118), and the occurrence of cerebral vasospasm-related events as a binary outcome, which was the primary endpoint of Clazosentan phase 3 trial.
We use covariate data of Clazosentan phase 3 trial as (current) trial data (the sample size is 220), and generate assignment results to the clazosentan 10 mg/hr group ($A=1$) or the placebo group ($A=0$) that differ from those of the original Clazosentan phase 3 trial by applying each of five assignment designs: SR, COVCAR \citep{ma2022CAR}, PROCAR1, PROCAR2 and PROCAR3. 
We set covariate features for COVCAR as $\phi (X)=(1,\mathrm{Age}, \mathbb{I}(\mathrm{WFNS}=\mathrm{I}),\mathbb{I}(\mathrm{WFNS}=\mathrm{II}))^\top$.
We train the prognostic score model $m^{(a)}(x;\alpha^{(a)})=\mathrm{expit}\{\alpha ^{(a)\top} \phi (x)\}$ for $a=0,1$ using historical data for PROCAR1--3. 
Then, $\alpha^{(1)}$ is estimated as $\widetilde \alpha^{(1)}=(-7.020, 0.100, -0.745, -0.004)^\top$ and $\alpha^{(0)}$ is estimated as $\widetilde \alpha^{(0)}=(-1.746, 0.057, -2.621, -1.041)^\top$. 
We set $m^{(a)}(x;\widetilde\alpha^{(a)})$ as $\widetilde m^{(a)}(x)$.
Figure \ref{fig: ana prog plot} shows boxplots of the between-group difference in the mean of $\widetilde{g}(X)$ (i.e., $\sum_{i=1}^n A_i \widetilde{g}(X_i)/\sum_{i=1}^n A_i - \sum_{i=1}^n (1-A_i)\widetilde{g}(X_i)/\sum_{i=1}^n(1-A_i)$) by changing the random seed and repeating the assignment for 500 iterations. 
Compared to existing assignment methods, our proposed CAR methods reduce $\widetilde{g}(X)$ imbalance.

\begin{figure}
    \centering
    \includegraphics[width=\linewidth]{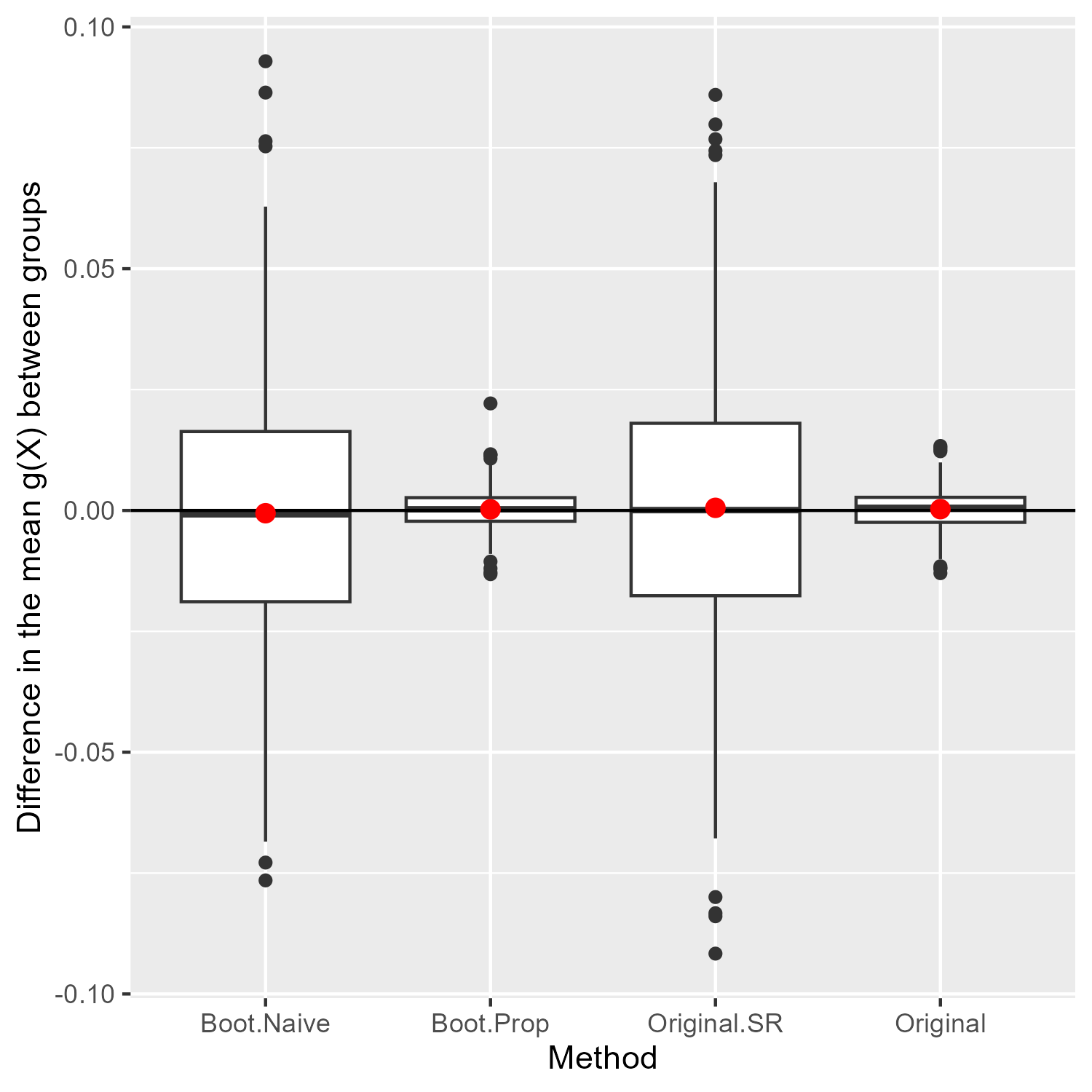}
    \caption{Results of applying bootstrap methods to Clazosentan phase 3 trial data after performing the assignment by PROCAR1. The two boxplots on the left show the between-group difference in the mean of $\widetilde{g}(X)$ in the bootstrap samples (i.e., $\sum_{i=1}^n A^*_{b,i} \widetilde{g}(X^*_{b,i})/\sum_{i=1}^n A^*_{b,i} - \sum_{i=1}^n (1-A^*_{b,i})\widetilde{g}(X^*_{b,i})/\sum_{i=1}^n(1-A^*_{b,i}), \ b=1,\ldots, 500=B$) after performing the assignment by PROCAR1; the leftmost is the naive bootstrap method (Boot.Naive), and the second from the left is our proposed bootstrap method (Boot.Prop). The two boxplots on the right show the between-group difference in the mean of $\widetilde{g}(X)$ in the original sample (i.e., $\sum_{i=1}^n A_i \widetilde{g}(X_i)/\sum_{i=1}^n A_i - \sum_{i=1}^n (1-A_i)\widetilde{g}(X_i)/\sum_{i=1}^n(1-A_i)$) by changing the random seed and repeating the assignment for 500 iterations; the second from the right is the distribution by simple randomization (Original.SR), and the rightmost is the distribution by PROCAR1 (Original). The red dots represent the mean of the between-group difference in the mean of $\widetilde{g}(X)$ over 500 iterations. The black horizontal line represents a between-group difference of zero.}
    \label{fig: ana boot prog plot}
\end{figure}

We also apply the naive and proposed bootstrap methods after performing the assignment by PROCAR1.
Figure \ref{fig: ana boot prog plot} evaluates whether the bootstrap methods reproduce the distribution induced by PROCAR1. 
The two boxplots on the left show the between-group difference in the mean of $\widetilde{g}(X)$ in the bootstrap samples (i.e., $\sum_{i=1}^n A^*_{b,i} \widetilde{g}(X^*_{b,i})/\sum_{i=1}^n A^*_{b,i} - \sum_{i=1}^n (1-A^*_{b,i})\widetilde{g}(X^*_{b,i})/\sum_{i=1}^n(1-A^*_{b,i}), \ b=1,\ldots, 500=B$) after performing the assignment by PROCAR1; the leftmost is the naive bootstrap method (Boot.Naive), and the second from the left is our proposed bootstrap method (Boot.Prop). 
The two boxplots on the right show the between-group difference in the mean of $\widetilde{g}(X)$ in the original sample (i.e., $\sum_{i=1}^n A_i \widetilde{g}(X_i)/\sum_{i=1}^n A_i - \sum_{i=1}^n (1-A_i)\widetilde{g}(X_i)/\sum_{i=1}^n(1-A_i)$) by changing the random seed and repeating the assignment for 500 iterations; the second from the right is the distribution by simple randomization (Original.SR), and the rightmost is the distribution by PROCAR1 (Original).
The proposed bootstrap method closely reproduces the distribution induced by PROCAR1, whereas the naive bootstrap method produces a distribution resembling that induced by SR.

\section{Concluding remarks}
\label{sec: discuss}
In this study, we made two proposals for CAR.
The first proposal is CAR using historical data.
The proposal is the random assignment design to minimize the imbalance in $g(X)$.
$g(X)$ is the score that integrates multidimensional covariates into a one-dimensional score while accounting for the magnitude of their associations with the outcome.
In simulation studies, our proposed CAR methods reduced the imbalance in $g(X)$, thereby improving the precision of ATE estimation. 
The second proposal is a bootstrap method for inference under CAR, which uses unequal-probability resampling to account for the assignment design.
In simulation studies, our proposed bootstrap method accurately assessed estimation uncertainty under our proposed CAR.
In data application, our proposed CAR methods achieved a better balance in sample size and $\widetilde g(X)$ than the CAR method of \cite{ma2022CAR}, and our proposed bootstrap method appropriately reproduced the distribution induced by our proposed CAR method.

As demonstrated in simulations, our proposed methods allow for efficient estimation even in the presence of covariate shift. 
However, in the presence of outcome shift, the prognostic score learned from historical data, and consequently $\widetilde{g}(x)$, will be misspecified for the current trial population. 
This may limit the efficiency gain, although consistency is preserved.
Thus, in practice, it is important to use historical data that are sufficiently similar to the current trial data.
Furthermore, there may be important covariates that are available in current trial data but were not collected in historical data, or the levels of important covariates that exist in current trial data may not exist in historical data (e.g., site).
In this case, it would be better to train the prognostic score without those covariates and then use an imbalance measure that combines the prognostic score with them.

Our proposed methods allow for efficient estimation and valid inference without adjustment at the analysis stage and can be easily extended to any type of outcome. 
In particular, they enable estimation of marginal hazard ratios using a Cox model with treatment assignment as the only covariate, thereby avoiding issues related to non-collapsibility, as well as estimation of marginal survival curves using the Kaplan-Meier estimator, which ensures boundedness and monotonicity.

\bibliographystyle{apalike}
\bibliography{PROCAR}

\end{document}